\documentclass[aps,pre,reprint,showpacs,amsmath,amssymb,superscriptaddress]{revtex4-1}

\usepackage{graphicx}
\usepackage{bm}
\graphicspath{ {./fig/} }



\begin{document}

\title{Accounting for adsorption and desorption in Lattice Boltzmann simulations}


\author{Maximilien Levesque}
\email{maximilien.levesque@gmail.com}
\affiliation{UPMC Univ Paris 06, UMR 7195, PECSA, F-75005, Paris, France}
\affiliation{CNRS, UMR 7195, PECSA, F-75005, Paris, France}
\affiliation{Department of Materials, University of Oxford, Parks Road, Oxford OX1 3PH, United Kingdom}

\author{Magali Duvail}
\affiliation{UPMC Univ Paris 06, UMR 7195, PECSA, F-75005, Paris, France}
\affiliation{CNRS, UMR 7195, PECSA, F-75005, Paris, France}
\affiliation{Institut de Chimie S\'eparative de Marcoule, ICSM, UMR 5257, CEA-CNRS-Université Montpellier 2-ENSCM, Site de Marcoule, BP 17171, F-30207 Bagnols-sur-C\`eze, France}

\author{Ignacio Pagonabarraga}
\affiliation{Departament de F\'isica Fonamental, Universitat de Barcelona, 08028
Barcelona, Spain}

\author{Daan Frenkel}
\affiliation{Department of Chemistry, University of Cambridge, Lensfield Road, CB2 1EW Cambridge, United Kingdom}

\author{Benjamin Rotenberg}
\affiliation{UPMC Univ Paris 06, UMR 7195, PECSA, F-75005, Paris, France}
\affiliation{CNRS, UMR 7195, PECSA, F-75005, Paris, France}

\begin{abstract}
We report a Lattice-Boltzmann scheme that accounts for adsorption
and desorption in the calculation of mesoscale dynamical properties
of tracers in media of arbitrary complexity. Lattice Boltzmann simulations
made it possible to solve numerically the coupled Navier-Stokes equations
of fluid dynamics and Nernst-Planck equations of electrokinetics in
complex, heterogeneous media. Associated to the moment propagation
scheme, it became possible to extract the effective diffusion and
dispersion coefficients of tracers, or solutes, of any charge, \emph{e.g.}
in porous media. Nevertheless, the dynamical properties of tracers
depend on the tracer-surface affinity, which is not purely electrostatic,
but also includes a species-specific contribution. In order to capture
this important feature, we introduce specific adsorption and
desorption processes in a Lattice-Boltzmann scheme through a modified
moment propagation algorithm, in which tracers may adsorb and desorb
from surfaces through kinetic reaction rates. The method is validated
on exact results for pure diffusion and diffusion-advection in Poiseuille
flows in a simple geometry. We finally illustrate the
importance of taking such processes into account on the time-dependent
diffusion coefficient in a more complex porous medium.
\end{abstract}
\maketitle

\section{Introduction}

The dynamical properties of fluids in heterogeneous
materials offers a great challenge and have implications in many 
technological and environmental contexts.
Inherently multi-scale in time and space, mesoscale properties such as
the diffusion or dispersion coefficient reflect the
nano-to-microscopic geometry of the media and the inter-atomic interactions
between flowing particles, or tracers, and surface atoms. Experimentally,
information about the microstructure of porous media can be extracted
from diffusion measurements by pulsed gradient spin echo nuclear magnetic
resonance (PGSE-NMR)~\cite{mitra_diffusion_1997,sen_time-dependent_2003,sen_time-dependent_2004}.
At short times, the dynamics of a pulse of tracers is connected to
the geometry of the porous medium at the pore scale. At longer times,
macroscale properties such as porosity and tortuosity come into play.
Theoretically, stochastic approaches have been an important support
to the understanding of the underlying phenomena~\cite{sen_time-dependent_2003,sen_time-dependent_2004,dudko_time-dependent_2005},
and have been used recently to show that adsorption and desorption
processes may strongly modify the short and long time dynamics of
the tracers~\cite{levesque_taylor_2012,levesque_note_2013}.

In numerous if not all practical situations involving particle diffusion
and advection, the carrier fluid is in contact with confining walls
where adsorption may occur. These processes depend on the chemical
nature of the solute, which explain why particles with the same charge
may diffuse in the same medium with different effective
diffusion coefficients. This species-dependent affinity is at the
heart of all chromatographic techniques used in analytic and separation
chemistry \cite{tallarek_study_1998}. It also plays a crucial role
in the dissemination of toxic or radioactive pollutants in the environment,
and conversely in remediation strategies. Recently, the great interest
for nanofluidic devices and for the transport in heterogeneous porous
media has also raised the issue of the relevance of models which do
not take into account these sorption processes. Moreover, it was recently
shown that stochastic resonance between these processes and some external
field may be of practical importance, e.g. for molecular sorting~\cite{alcor_molecular_2004,levesque_taylor_2012}.

At the mesoscale, the dynamics of particles in a fluid can be described
by the continuity equation
\begin{equation}
\partial_{t}\rho\left(\mathbf{r},t\right)=-\mathbf{\nabla}\cdot\mathbf{J}\left(\mathbf{r},t\right),\label{eq:advection-diffusion}
\end{equation}
where $\rho(\mathbf{r},t)$ is the one-particle density at position
$\mathbf{r}$ and time $t$, $\partial_t \equiv \partial / \partial t$ and $\mathbf{J}$ is the particle flux,
which is a function of the velocity field of the carrier fluid, the
bulk diffusion coefficient $D_{b}$ of the particles and, if any,
their charge and the local electric field arising from their environment.
If adsorption is taken into account, solid-liquid interfaces
located at $\mathbf{r}$ have a surface concentration $\Gamma\left(\mathbf{r},t\right)$
(length$^{-2}$) that evolves with time according to:
\begin{equation}
\partial_{t}\Gamma\left(\mathbf{r},t\right)=-k_{d}\Gamma\left(\mathbf{r},t\right)+k_{a}\rho\left(\mathbf{r},t\right),\label{eq:surface_concentration}
\end{equation}
where $k_{a}$ (length$\cdot$time$^{-1}$) and $k_{d}$ (time$^{-1}$)
are kinetic adsorption and desorption rates. For molecules, the rates can vary widely.
As an example, the dissociation rate of DNA double strands on a surface grafted with single-strand DNA ranges
from $10^{-5}$ to $10^{-3}$~s$^{-1}$ for a few tens of base pairs~\cite{gunnarsson_DNAsorption_2007}.
Their adsorption rate can be adjusted by changing the grafting density.
Finally, we assume that the tracers (the solutes) neither diffuse into the solid phase 
(even though that process can easily be accounted for using our algorithm) nor dissolve the surfaces~\cite{Ladd_erosion_PRE_2002}.

The time-dependent diffusion coefficient $D(t)$ and the dispersion
coefficient $K$ of the tracers can be investigated by following the spreading
of a tracer pulse in the fluid. This would amount to solving Eqs.~\ref{eq:advection-diffusion}
and \ref{eq:surface_concentration}, \emph{e.g.} with a finite element
method, for all possible initial conditions, which is computationally
intractable for complex systems such as heterogeneous porous media. An
alternative is to deduce $K$ and $D(t)$ from the tracer velocity
auto-correlation function (VACF) following~\cite{lowe_super_1995,Lowe_Frenkel_1996_EPL1,Capuani_Frenkel_2003_EPL2,rotenberg_faraday_2010}:
\begin{eqnarray}
D_{\gamma}\left(t\right) & = & \int_{0}^{t}Z_{\gamma}\left(t^{\prime}\right)\mbox{d}t^{\prime},\label{eq:D_t_from_vacf}\\
K_{\gamma} & = & \int_{0}^{\infty}\left(Z_{\gamma}\left(t\right)-Z_{\gamma}\left(\infty\right)\right)\mbox{d}t,\label{eq:K_from_vacf}
\end{eqnarray}
where the VACF in the direction $\gamma\in\left\{ x,y,z\right\}$
is
\begin{equation}
Z_{\gamma}\left(t\right)=\left\langle
v_{\gamma}(0)v_{\gamma}\left(t\right)\right\rangle.
\label{eq:def_vacf}
\end{equation}
At long times $Z_{\gamma}\left(\infty\right)=\bar{v}_{\gamma}^{2}$
with $\bar{v}_{\gamma}$ the average velocity of the flow.
The issue of averaging over initial conditions in Eq.~\ref{eq:def_vacf}
can be handled elegantly and efficiently using the moment propagation
method~\cite{Lowe_Frenkel_1996_EPL1,Hoef_Frenkel_1990_EPL3,Merks_2002_EPL4},
which was recently extended to charged tracers~\cite{Rotenberg_EPL_2008,pagonabarraga_PCCP_2010,rotenberg_faraday_2010,wang2009electrokinetic,Wang2010electrokinetic,Wang2010electroosmosis}.

In order to compute the VACF of the tracers from Eq.~\ref{eq:def_vacf},
one has to keep track of their velocity.
For this purpose, we use the underlying dynamics of the fluid given by Eq.~\ref{eq:advection-diffusion}, which does not rely on the velocity of individual particles, but on the one-particle solvent density.
Moreover, the simulation of heterogeneous multi-scale media 
requires a numerically efficient method. 
The lattice-Boltzmann (LB) method \cite{frisch_lattice-gas_1986,mcnamara_use_1988,ladd_theoretical_1994,ladd_numerical_1994,succi_lb,aidun_lattice-boltzmann_review_2010,chen_lattice_1998}
offers a convenient framework to deal with such situations.
In the LB approach, the fundamental quantity is a one-particle
velocity distribution function $f_{i}\left(\mathbf{r},t\right)$ that
describes the density of particles with velocity $\mathbf{c}_{i}$,
typically discretized over 19 values for three-dimensional LB, at
a node $\mathbf{r}$, either fluid or solid, of a lattice of spacing
$\Delta x$, and at a time $t$ discretized by steps of $\Delta t$.
The dynamics of the fluid are governed by transition probabilities
of a particle moving in the fluid from one node to the neighboring
ones: 
\begin{equation}
f_{i}\left(\mathbf{r}+\mathbf{c}_{i}\Delta t,t+\Delta t\right)=f_{i}\left(\mathbf{r},t\right)+\Delta_{i}\left(\mathbf{r},t\right),\label{eq:lb_evolution}
\end{equation}
where $\Delta_{i}$, the so-called collision operator, is the change
in $f_{i}$ due to collisions at lattice nodes. This LB equation recovers the fluid dynamics of a liquid and the moments of the distribution function are related to the relevant hydrodynamic variables. The reader is
referred to Refs.~\cite{succi_lb} and~\cite{chen_lattice_1998}
for reviews of the method.

\section{Algorithm}
In order to compute the dynamical properties of tracers evolving in
a fluid described by the LB algorithm, \emph{i.e.} to solve Eqs.~\ref{eq:D_t_from_vacf}--\ref{eq:def_vacf},
we use the moment propagation (MP) method~\cite{lowe_super_1995,Lowe_Frenkel_1996_EPL1}.
Other methods could have been used, such as the numerical resolution of the macroscopic equations or Brownian Dynamics to simulate the random walk of tracers biased by the LB flow~\cite{maier_pore-scale_1998}. The latter method has often been successfully used, e.g. by Boek and Venturoli~\cite{Boek20102305}. Nevertheless, the MP method offers many advantages. First, MP relies on the same
ground as LB. It is based on the propagation of a position and velocity distribution function~\cite{Lowe_Frenkel_1996_EPL1,Hoef_Frenkel_1990_EPL3}, therefore offering an elegant unified approach. Secondly, MP allows for the propagation of \emph{any} moment of the distribution function ~$f_{i}\left(\mathbf{r},t\right)$, which offers great opportunities. For example, Lowe, Frenkel and van~der~Hoef exploited these higher moments to compute self-dynamic structure factors~\cite{lowe_1997_EPL6}. The LB-MP method has been thoroughly validated by Merks for low P\'eclet and Reynolds numbers~\cite{Merks_2002_EPL4}.

In the moment propagation algorithm, any quantity $P\left(\mathbf{r},t\right)$ can be propagated 
between fluid nodes. This quantity will be modified by adsorption and desorption processes. In their absence, $P\left(\mathbf{r},t+\Delta t\right)=P^{\star}\left(\mathbf{r},t+\Delta t\right)$ with:

\begin{eqnarray}
P^{\star}\left(\mathbf{r},t+\Delta t\right) & = & \sum_{i}\left[P\left(\mathbf{r}-\mathbf{c}_{i}\Delta t,t\right)p_{i}\left(\mathbf{r}-\mathbf{c}_{i}\Delta t,t\right)\right]\nonumber \\
 &  & +P\left(\mathbf{r},t\right)\left(1-\sum_{i}p_{i}\left(\mathbf{r},t\right)\right),\label{eq:propagated_quantity}
\end{eqnarray}
where the first sum runs over all discrete velocities connecting adjacent
nodes. The probability of leaving node $\mathbf{r}$ along the direction
$\mathbf{c}_{i}$ is noted $p_{i}\left(\mathbf{r},t\right)$. The
last term in Eq.~\ref{eq:propagated_quantity} represents the fraction
of particles that did not move from $\mathbf{r}$ at the previous
time-step. The expression for $p_{i}$, which is central in the algorithm,
depends on the nature of the tracers. It is given by: 
\begin{equation}
p_{i}\left(\mathbf{r},t\right)=\frac{f_{i}\left(\mathbf{r},t\right)}{\rho\left(\mathbf{r},t\right)}-\omega_{i}+\frac{\omega_{i}\lambda}{2}Q,
\end{equation}
where the first two terms account for advection and are obtained by
coupling the tracer dynamics to that of the fluid evolving according
to the LB scheme. The weigths $\omega_{i}$ are constants depending
upon the underlying LB lattice. The last term describes diffusive
mass transfers. The dimensionless parameter $\lambda$ determines
the bulk diffusion coefficient $D_{b}=\lambda c_{s}^{2}\Delta t/4$,
with $c_{s}=\sqrt{k_{B}T/m}$ the sound velocity in the fluid. 
It also determines the mobility of tracers under the influence of chemical potential
gradients (including the electrostatic contribution) which are accounted for in
the $Q$ term as described in Ref.~\cite{Rotenberg_EPL_2008}.
For neutral tracers, $Q=1$.

We now introduce a new propagation scheme in order to account for
adsorption and desorption at the solid-liquid interface. While Eq.~\ref{eq:propagated_quantity}
still holds for nodes $\mathbf{r}$ which are in the fluid but not
at the interface, for the fluid interfacial nodes we define a
new propagated quantity $P_{\textrm{ads}}\left(\mathbf{r},t\right)$ 
associated with adsorbed particles:
\begin{eqnarray}
P_{\textrm{\mbox{ads}}}\left(\mathbf{r},t+\Delta t\right) & = & P\left(\mathbf{r},t\right)p_{a}+P_{\textrm{\mbox{ads}}}\left(\mathbf{r},t\right)\left(1-p_{d}\right),\label{eq:Pads}
\end{eqnarray}
where $p_{a}=k_{a}\Delta t/\Delta x$ is the probability for a tracer
lying at an interface to adsorb, and $p_{d}=k_{d}\Delta t$ is
the probability for an adsorbed, immobile tracer to desorb. There is no restriction
in the definition of $k_{a}$ and $k_{d}$ so that they may depend
on geometrical considerations and on the local tracer density. Finally,
the evolution of the propagated quantity associated with free tracers
now includes a term accounting for the desorption of adsorbed particles:
\begin{eqnarray}
P\left(\mathbf{r},t+\Delta t\right) & = & P^{\star}\left(\mathbf{r},t+\Delta t\right)+P_{\textrm{\mbox{ads}}}\left(\mathbf{r},t\right)p_{d}.\label{eq:newP}
\end{eqnarray}
where $P^{\star}$ is still given by Eq.~\ref{eq:propagated_quantity}.
In order to compute the VACF of the tracers, one propagates as $P$ 
the probability to arrive at position~\textbf{$\mathbf{r}$} at time $t$, 
weighted by the initial velocity of tracers. 
Thus one needs to initialize, for each direction $\gamma$, 
a propagated quantity according to the Maxwell-Boltzmann distribution. 
The Boltzmann weights for solid ($\mathcal{S}$), fluid ($\mathcal{F}$)
and interfacial ($\mathcal{I}\subset{\cal F}$) nodes read:
\begin{equation}
\begin{cases}
0 & \mbox{for }\mathbf{r}\in{\cal S},\\
e^{-\beta\mu^{ex}\left(\mathbf{r}\right)}/\mathcal{Z} & \mbox{for }\mathbf{r}\in{\cal F\setminus I},\\
e^{-\beta\mu^{ex}\left(\mathbf{r}\right)}\left(1+e^{-\beta\Delta\mu^{ads}\left(\mathbf{r}\right)}\right)/\mathcal{Z} & \mbox{for }\mathbf{r}\in{\cal I},
\end{cases}
\end{equation}
where $e^{-\beta\Delta\mu^{ads}\left(\mathbf{r}\right)}=k_{a}/\left(k_{d}\Delta x\right)$
corresponds to the sorption free energy for interfacial tracers,
$\beta\equiv1/k_{B}T$ with $k_{B}$ the Boltzmann's constant and
$T$ the temperature,
and $\mathcal{Z}$ is the partition function of the tracers. The
excess chemical potential $\mu^{ex}\left(\mathbf{r}\right)$
includes in the case of tracers with charge $q$ a mean-field
electrostatic contribution $q\psi\left(\mathbf{r}\right)$
with $\psi$ the local electrostatic potential.
The VACF is then simply given as in the no sorption
case~\cite{Rotenberg_EPL_2008} by:
\begin{equation}
Z_{\gamma}(t)=\sum_{\mathbf{r}}P\left(\mathbf{r},t\right)\times\left(\sum_{i}p_{i}\left(\mathbf{r},t\right)c_{i\gamma}\right).
\end{equation}

\section{Validation}

In order to validate this new scheme, we compare numerical results and exact theoretical solutions of Eqs.~\ref{eq:advection-diffusion} and \ref{eq:surface_concentration}. This allows to assess the validity of our method independently of experimental results. We consider the
diffusion and dispersion of tracers in a slit pore, \emph{i.e.} between
two walls at positions $x=0$ and $x=L$. The time-dependent diffusion
coefficient $D\left(t\right)$ in the direction normal to the wall
is given by~\cite{levesque_note_2013}:

\begin{eqnarray}
\frac{D(t)}{D_{b}} & = & \frac{1}{\left(2k_{a}+k_{d}L\right)}\times\mathcal{L}^{-1}\left[\frac{k_{d}L}{\chi^{2}}\right.\nonumber \\
 & - & \left.\frac{2k_{d}\left(k_{d}+s\right)\sinh\kappa}{\chi^{3}\left(\left(k_{d}+s\right)\cosh\kappa+k_{a}\chi\sinh\kappa\right)}\right],\label{eq:exact_D}
\end{eqnarray}
where $s$ is the Laplace conjugate of time $t$, $\mathcal{L}^{-1}$
is the inverse Laplace transform, $\chi=\sqrt{s/D_{b}}$ and $\kappa=\chi L/2$.

\begin{figure}
\includegraphics[width=\columnwidth]{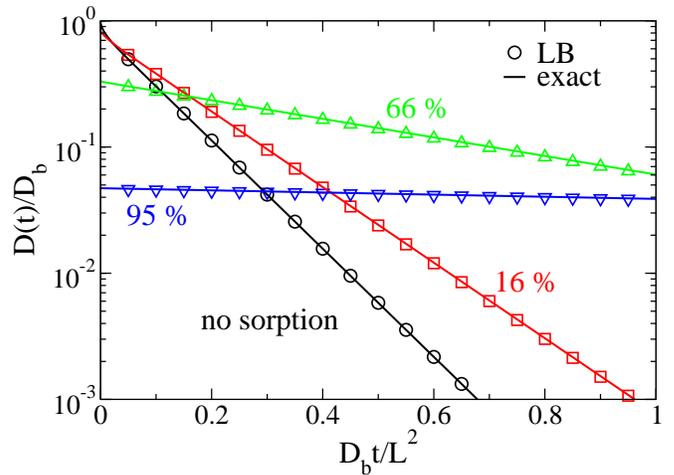}
\caption{(Color online) The time-dependent diffusion coefficient, $D(t)$,
normalized by the bulk diffusion coefficient, $D_{b}$, of neutral
tracers in a slit pore, as extracted from Lattice-Boltzmann simulations
using our new scheme (symbols) and from the reference exact solution
of Eq.~\ref{eq:exact_D}~(lines). Several fractions of adsorbed
tracers, or sorption strength, $f_a$ defined in Eq.~\ref{eq:fa}, are presented: black circles, 0\%;
red squares, 16\%; green upward triangles, 66\%; and blue downward triangles, 95\%.\label{fig:Doft}}
\end{figure}

In Fig.~\ref{fig:Doft}, we compare the exact time-dependent diffusion
coefficient $D(t)$ of Eq.~\ref{eq:exact_D} with the one extracted from Lattice-Boltzmann simulations for
different sorption strength. This last quantity is defined by the fraction of adsorbed
tracers. Unless otherwise stated, all simulations are performed within a slit pore of width
$L=100$~$\Delta x$ and a bulk diffusion coefficient $D_{b}=10^{-2}$~$\Delta x^{2}/\Delta t$.
These values are chosen very conservative, since the LB method is known to be efficient even in narrow slits (even for $L<10$~$\Delta x$) and for a wide range of magnitudes in the diffusion coefficients~\cite{Merks_2002_EPL4,ladd_numerical_1994}. $L$ and $D_b$ therefore account for a negligible part of the difference with exact results. We can thus purposely assess the effect of the new algorithm only.
In Fig.~\ref{fig:Doft}, we report time-dependent diffusion coefficients calculated by our method for a fixed sorption rate $k_a=10^{-1}~\Delta x/\Delta t$
and decreasing desorption rates $k_d \Delta t=10^{-2}$, $10^{-3}$ and $10^{-4}$ resulting in an increasing fraction of adsorbed tracers $f_a$ of approximately 16~\%, 66~\% and 95~\%. This fraction is given by~\cite{levesque_note_2013}:
\begin{equation}\label{eq:fa}
f_a=\left(1+\frac{k_d L}{2k_a}\right)^{-1}.
\end{equation}
Excellent agreement is found between exact solutions and our numerical results for all fractions of adsorbed tracers, \emph{i.e.} all "sorption strengths".

At the initial time, $D(t=0)$
is given by the fraction of mobile tracers. At intermediate times,
sorption/desorption processes significantly decrease the slope of $D(t)$,
as the partial immobilization at the surface slows down the exploration of the pore.
We have shown recently that for this range of parameters, this slope is given by $k_d/(1+k_aL/2D_b)$~\cite{levesque_note_2013}.
In this illustration of diffusion between two walls, the confinement
is total so that for sufficiently long times, the effective diffusion
coefficient decreases exponentially with time and tends to zero.

\begin{figure}
\includegraphics[width=8cm]{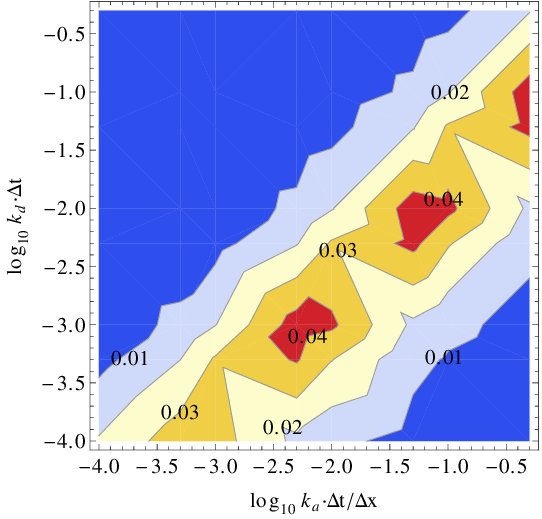}
\includegraphics[width=8cm]{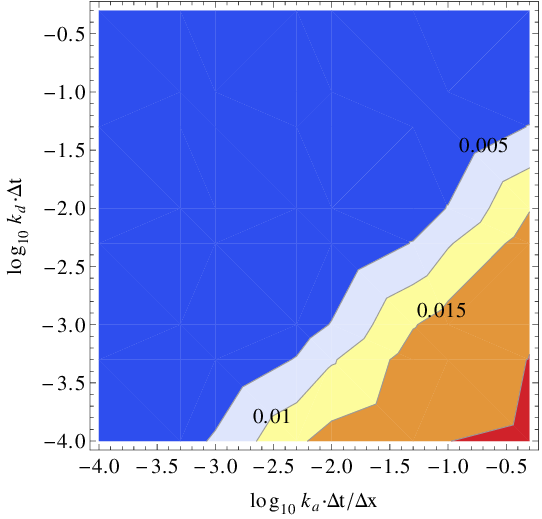}
\caption{(Color online) Contour plots of the relative error of our Lattice
Boltzmann scheme with respect to exact results, given by Eq.~\ref{eq:exact_D},
on the slope (top) and origin (bottom) of the linear regression of
the time-dependent diffusion coefficient of neutral tracers in a slit
pore, as a function of the adsorption and desorption rates $k_{a}$
and $k_{d}$.\label{fig:rel_err_Doft}}
\end{figure}

In order to quantify the error on $D(t)$ with respect to the exact result
of Eq.~\ref{eq:exact_D}, we plot in figure~\ref{fig:rel_err_Doft}
the relative error on the slope and origin of the linear fit of $\log_{10}D(t)$
for long times, \emph{i.e.} for $D_{b}t/L^{2}>0.5$, as a function of $k_{a}$ and $k_{d}$.
In the whole range of $k_{a}$ and $k_{d}$, the relative errors on the
slope and origin remain under 5~\% and 2~\%, respectively, which
is highly satisfactory.

The effect of a pressure gradient has also been studied on the same 
system. The resulting Poiseuille flow 
induces Taylor-Aris dispersion~\cite{taylor_dispersion_1953,aris_dispersion_1956}
of the tracers with a dispersion coefficient $K$, which is known
exactly in the presence of adsorption and desorption in the simple slit
geometry~\cite{levesque_taylor_2012}: 
\begin{eqnarray}
\frac{K}{D_{b}} & = &
1+P_{e}^{2}\left[\frac{102y^{2}+18y+1}{210\left(1+2y\right)^{3}}
+\frac{D_{b}}{L^2k_d}\frac{2y}{\left(1+2y\right)^{3}}\right], \label{eq:exact_K}
\end{eqnarray}
where $P_{e}=L\bar{v}/D_{b}$ is the P\'eclet number
and $y=k_a/k_dL$.

\begin{figure}
\includegraphics[width=\columnwidth]{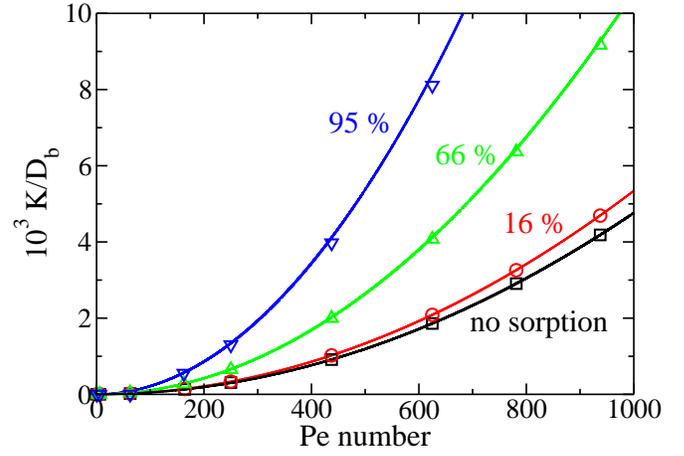}
\caption{(Color online) Dispersion coefficient of neutral tracers in a slit
pore in the direction of the flux, normalized by the bulk diffusion
coefficient, as a function of the P\'eclet number, as extracted from
our Lattice-Boltzmann scheme (symbols) and from the exact results~(lines).
Several fractions of adsorbed tracers, or sorption strength, $f_a$ defined in Eq.~\ref{eq:fa}, are
presented: black squares, 0\%; red circles, 16\%; green upward triangles, 66\%;
and blue downward triangles, 95\%.\label{fig:dispersion}}
\end{figure}

In Fig.~\ref{fig:dispersion}, we compare the dispersion coefficient
as calculated by LB with the exact results of Eq.~\ref{eq:exact_K}
for various sorption strengths, as a function of the P\'eclet number.
Adsorption significantly increases the dispersion, as it slows down
part of the tracers. The agreement between our scheme and exact results
is excellent, even for strong adsorption and P\'eclet numbers above
100.

\begin{figure}
\begin{centering}
\includegraphics[width=0.5\columnwidth]{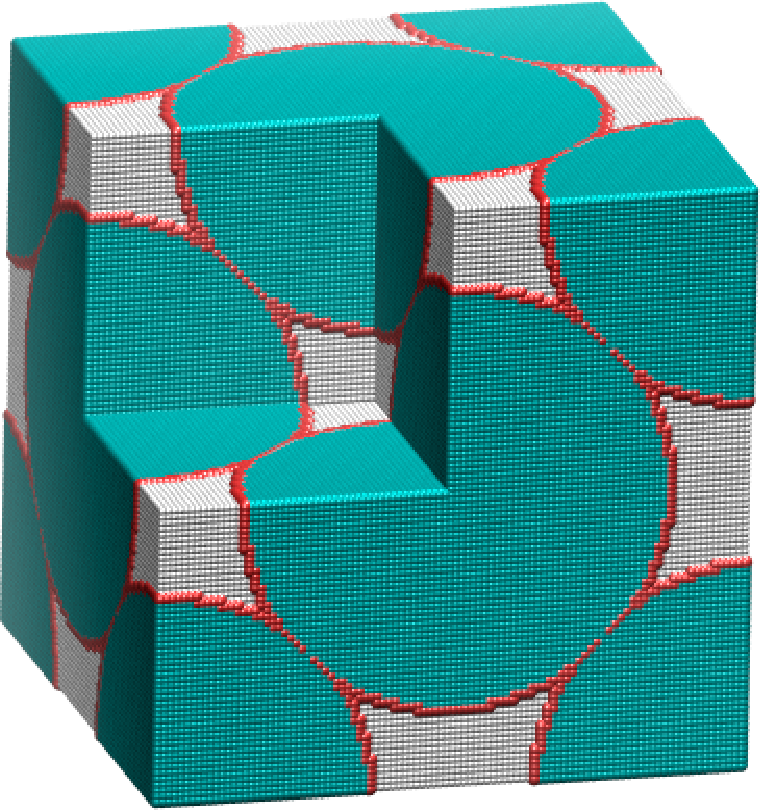}
\par\end{centering}
\caption{(Color online) Perspective view of a unit cell of a face-centered
cubic packing of spheres, of lattice parameter 100~$\Delta x$. Solid
nodes are colored in blue. Interfacial fluid nodes, where adsorption
processes may occur, are colored in red. Non-interfacial fluid nodes
are in white.\label{fig:bcc}}
\end{figure}

As mentioned above, the time-dependence of the diffusion coefficient
is a signature of the intrinsic geometric properties of a porous medium.
The simplest model of such media consists of a compact face
centered cubic (fcc) lattice of spheres of radius $R$~\cite{ladd_dissolution_2001}. The porosity,
\emph{i.e.} the fraction of empty (or fluid) space, is $1-\pi/\left(3\sqrt{2}\right)\approx26$\%.
The unit cell contains 4 octahedral cavities of radius $\approx0.41R$
connected by 8 smaller tetrahedral cavities of radius $\approx0.22R$
by a small channels of radius $\approx0.15R$. This fcc lattice
is illustrated in Fig.~\ref{fig:bcc} for a lattice parameter $L=100\Delta x$.

\begin{figure}
\includegraphics[width=\columnwidth]{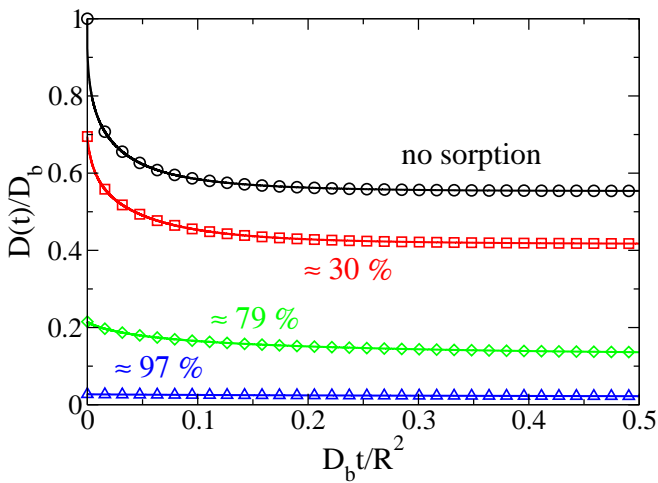}
\caption{(Color online) Diffusion coefficient $D(t)$ of a neutral tracer,
normalized by the bulk diffusion coefficient $D_{b}$ in the face-centered
cubic packing of spheres with radius $R$ illustrated in Fig.~\ref{fig:bcc},
as a function of the reduced time. Several fractions of adsorbed
tracers, or sorption strength, $f_a$ defined in Eq.~\ref{eq:fa}, are presented: black circles, 0\%,
\emph{i.e.}, without adsorption; red squares, 30\%; green upward triangles, 79\%;
and blue downward triangles, 97\%.\label{fig:diffusion_bcc}}
\end{figure}

We report the time-dependent diffusion coefficient for this model porous medium 
in Fig.~\ref{fig:diffusion_bcc}. At $t=0$, the diffusion
coefficient is again given by the fraction of free particles times
their bulk diffusion coefficient. After a reduced time 
$D_bt/R^2=0.5$, tracers
have explored the whole porosity and the diffusion coefficient tends
toward the effective diffusion coefficient. The time
dependence is strongly influenced by adsorption/desorption, as
in the slit pore case. 
It is thus essential to consider these phenomena when interpreting
experimental measurement of effective and time-dependent diffusion coefficients.

\section{Conclusion}
In summary, we have proposed a new scheme that accounts for adsorption
and desorption in a generic Lattice-Boltzmann scheme, 
allowing for the calculation of mesoscale dynamical properties 
of tracers in media of arbitrary complexity. These processes
are modelled by kinetic rates of adsorption and desorption taking place
at interfacial, fluid nodes. The algorithm has been validated 
over a wide range of adsorption and desorption rates and P\'eclet
numbers in the slit pore geometry where exact results are available. 
Finally, we have shown on a more complex porous medium that adsorption
and desorption processes may not be neglected, as they strongly modify
the short, intermediate and long time behaviors of the diffusion coefficient
as well as the dispersion coefficient.
In turn, this demonstrates that neglecting interactions with the surface
in the interpretation of $D(t)$ as a probe of the geometry of the porous
medium, as measured experimentally \emph{e.g.} by PSGE-NMR, 
may lead to incorrect conclusions.
This scheme may now be used in two ways. First, one could predict 
the effective diffusion coefficient in complex heterogeneous
media for species with known adsorption and desorption rates 
(from experiments or molecular simulations). Conversely, from reference
measurements of the time-dependent diffusion coefficient in controlled
geometries, one could extract the adsorption and desorption rates $k_{a}$ and $k_{d}$.
Moreover, in the case of diffusion in the solid stationary phase, the method would allow
 us to relate the relevant diffusion constant to the shape of the elution profile.

While the present method is very general and also applies in principle to the case of
irreversible adsorption ($k_d = 0$, i.e. $p_d= 0$ in Eq.~\ref{eq:newP}), such a situation
is of interest only outside of equilibrium. Indeed, in that case at equilibrium all the solute
is adsorbed on the surface and its VACF corresponds to the sorbed species only. The propagation
scheme (Eqs.~\ref{eq:propagated_quantity}-\ref{eq:newP}) could nevertheless be used to investigate
irreversible adsorption out of equilibrium by considering the density as the propagated quantity $P$
(instead of the one described here to compute the VACF), as was done e.g. by Warren to simulate electrokinetic
phenomena~\cite{warren_electroviscous_1997}. As an example of practical application where (possibly irreversible) sorption
is coupled to electrokinetic phenomena, we can for example mention the case of ion adsorption
onto charged minerals such as clays.

\begin{acknowledgments}
BR and ML acknowledge financial support from the French Agence Nationale
de la Recherche under grant ANR-09-SYSC-012.  MD acknowledges the French Agence Nationale pour la Gestion des D\'echets Radioactifs (ANDRA) for financial support. IP acknowledges Spanish MINECO (Project No. FIS2011-22603) and DURSI (SGR2009-634) for financial support. DF acknowledges a Wolfson Merit Award of the Royal Society of London.
\end{acknowledgments}

%

\end{document}